\documentstyle[twoside,fleqn,npb,epsfig]{article}
\newcommand{\AmS}{{\protect\the\textfont2
  A\kern-.1667em\lower.5ex\hbox{M}\kern-.125emS}}
\title{Running Coupling BFKL Equation and Deep Inelastic Scattering.}

\author{R.S. Thorne\address{Jesus College and Theoretical Physics,
        University of Oxford, \\
        Oxford, Oxon, OX1 3DW, U.K.}}

\begin{document}

\begin{abstract}
I examine the form of the solution of the BFKL equation with running coupling 
relevant for deep inelastic scattering. The evolution of structure
functions is precisely determined and well described by an effective
coupling $\sim 1/(\beta_0(\ln (Q^2/\Lambda^2)+3.6(\alpha_s(Q^2)
\ln(1/x))^{1/2})$ (until extremely small $x$). Corrections to the LO 
equation are relatively small, and the perturbative expansion is stable.
Comparison to data via a global fit is very successful. 
\end{abstract}

\maketitle

Small $x$ physics has recently been a particular area of both experimental and
theoretical interest. The first  
data on structure functions at very small values of $x$ (down to $x=10^{-5}$)
obtained by HERA [1,2] have themselves been
enough to make this an active topic. However, the fact that
the splitting
functions and coefficient functions required for the construction of structure
functions have expansions containing terms 
$\alpha^n_s(\mu^2)\ln^{n-1}(1/x)$
has added extra impetus,
implying that at small $x$  
one may have to account for leading terms
in $\ln(1/x)$ at high orders in $\alpha_s$, rather
than just expand naively in powers of $\alpha_s$.    

\section{The Running Coupling BFKL Equation.}

The BFKL equation [3] makes it possible to take account of the most leading 
$\ln(1/x)$ terms at each order in $\alpha_s$, since it is 
the four-point gluon Green's function containing all the
information about the most important small-$x$ behaviour. 
At leading order, with fixed $\alpha_s$, the BFKL equation is scale
invariant, and has simple eigenfunction $\sim (k^2)^{\gamma}$, with
eigenvalues $\chi^0(\gamma)$, where $\chi^0(\gamma)$ is the Mellin
transformation of the kernel. 
Transforming back to $x$ and $k^2$ space the leading small $x$ behaviour is 
driven by the saddle-point at $\gamma=1/2$, leading to small-$x$ behaviour of
the form $x^{-\lambda}$, 
where $\lambda=\bar \alpha_s\chi^0(1/2)=\bar\alpha_s 4\ln(2)$ ($\bar\alpha_s=
3\alpha_s/\pi$), i.e. if 
$\alpha_s \approx 0.2$, $\lambda \sim 0.5$. However, this result is
potentially subject to large corrections. In particular
the equation predicts large diffusion -- within the gluon Green's
function the average virtuality $\sim k^2$ but the spread of important
values has a width $\propto (\alpha_s\ln(1/x))^{1/2}$. Hence, there is
considerable influence from both the infrared and ultraviolet regions
of virtuality at small $x$. If the coupling is allowed to run this
will clearly be very important. 

Recently the NLO BFKL equation became available [4,5]. 
Not only does this provide information about the scale-invariant NLO
corrections, but the running of the QCD coupling now becomes impossible to
ignore -- the NLO part of the kernel contains a term $\propto \beta_0
\alpha^2_s(\mu^2) \ln(k^2/\mu^2)$, where $\mu$ is the renormalization scale, 
and this is absorbed into
the renormalization group improved coupling, $\alpha_s(\mu^2) -\beta_0
\alpha^2_s(\mu^2) \ln(k^2/\mu^2)\to \alpha_s(k^2)$. Having done this,
a simple way to consider the effect of the NLO corrections
to the BFKL kernel is to consider its action on the LO eigenfunctions 
$(k^2)^{\gamma}$. This results in a NLO {\it eigenvalue} of the form 
$\bar\alpha_s(k^2)\chi^0(\gamma) -\bar \alpha^2_s(k^2)\chi^1(\gamma)$, which
for $\gamma=1/2$
leads to an intercept of $2.8\bar\alpha_s(k^2)(1-6.5\bar\alpha_s(k^2))$. This
is clearly a disaster, implying a very unconvergent expansion for the
intercept. 

However, this simple approach requires modification. The {\it
eigenvalue} is not a real eigenvalue since it depends on $k^2$
-- the running coupling has broken scale invariance, 
changing the whole form of the BFKL equation. Even for the LO BFKL
equation, the $\ln(k^2/\Lambda^2)$ associated with a running coupling
turns the $\gamma$-space
BFKL equation into a differential equation. This can formally
be solved, and then inverted back to $k^2$-space. 
At leading twist, the solution factorizes into a $k^2$-dependent part
$g(k^2,N)$ and a $Q_0^2$-dependent part [6]. The latter is 
ambiguous since
the $\gamma$-dependent integrand has a cut along the contour of
integration of the inverse transformation. This is because the growth
of the coupling in the infrared coupled with the
strong infrared diffusion leads to an
infrared divergence in the BFKL equation, and hence a renormalon
contribution. Therefore the running coupling equation has no
prediction for the input for the gluon, this being sensitive to
nonperturbative physics. 
Conversely, $g(k^2,N)$ is completely
well-defined, being insensitive to
infrared diffusion. However, it is influenced the ultraviolet 
diffusion, and  hence by scales greater than $k^2$, where the coupling
is weak. 

Calculating $d g(k^2,N)/d\ln(k^2)$ one extracts an
unambiguous anomalous dimension $\Gamma(N,\alpha_s(k^2))$ governing the
evolution of the gluon. Considering for the moment the use of just the
LO kernel, $\Gamma(N, \alpha_s(k^2))$ turns out to be the usual
LO BFKL anomalous dimension $\gamma^0_{gg}(\bar\alpha_s/N)$, with
$\alpha_s$ evaluated at $k^2$, plus an infinite series of
corrections which are a power
series in $\beta_0\alpha_s(k^2)$ compared to the LO. Transforming to
$x$ space one finds that each term in the series behaves like
$x^{-\lambda(k^2)}$ as $x\to 0$, but with accompanying powers of
$\alpha_s(k^2) \ln(1/x)$ which grow as the power of $\beta_0
\alpha_s(k^2)$ grows. Hence a resummation of running coupling dependent
terms is necessary. The series is too complicated to sum exactly, 
so some prescription must be used. I choose the
BLM prescription [7], which adjusts the scale of the coupling in the LO
expression to $\tilde k^2$ so that using $\alpha_s^{eff}\equiv 
\alpha_s(\tilde k^2) =
\alpha_s(k^2) -\beta_0 \ln(\tilde k^2/k^2)\alpha_s^2(k^2)+ \cdots$ in
the LO splitting function one generates precisely the NLO
$\beta_0$-dependent correction to the splitting function. Choosing
$\tilde k^2$ in this manner one can find an exact expression for the
scale choice in the effective coupling, and in the small $x$ limit it
is $\ln(\tilde k^2/\Lambda^2) =\ln(k^2/\Lambda^2) +
3.6(\alpha_s(k^2)\ln(1/x))^{1/2}$, (see [8] for details), which is entirely
consistent with influence from the diffusion into the ultraviolet.
The effective coupling is shown in fig. 1.  

It is possible to check whether this prescription is sensible. If the
summation of the complete series is generated by this effective
scale then the whole series is given by using $\alpha_s(\tilde k^2) =
\alpha_s(k^2) -\beta_0 \ln(\tilde k^2/k^2)\alpha_s^2(k^2)+ \cdots$ in
the LO splitting function. This can be checked explicitly for the
${\cal O}((\beta_0\alpha_s(k^2))^2)$ term. The agreement between the
term generated in this manner and that calculated explicitly is 
excellent until very small $x$ indeed, i.e., the
${\cal O}((\beta_0\alpha_s(k^2))^2)$ term leads to a modification to
the scale of $\ln(\tilde k^2/\Lambda^2) =\ln(k^2/\Lambda^2) +
3.6(\alpha_s(k^2)\ln(1/x))^{1/2}-1.2\beta_0\alpha_s(k^2)
(\alpha_s(k^2)\ln(1/x))^2$. Checks at higher order 
in $\beta_0\alpha_s(k^2)$ give similar
results. 

\vskip -0.1in
\setlength{\unitlength}{0.7mm}
\begin{figure}[htb]
\begin{picture}(120,85)(-15,8)
\mbox{\epsfxsize5.0cm\epsffile{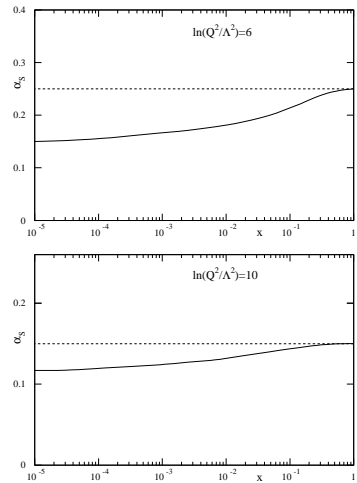}}
\end{picture}
\vskip -0.1in
\caption{$\alpha_s^{eff}$ as a function of $x$ compared to the constant values
for two choices of $\ln(Q^2/\Lambda^2)$.}
\label{fig:coupling}
\end{figure}
\vskip -0.2in

One can also solve for $\Gamma(N, \alpha_s(k^2))$ numerically and
compare with the transformed splitting function using the
effective scale. Due to a zero in $g(k^2, N)$ (first noticed in the context of
a resummed kernel in [9]), the anomalous dimension has a leading
pole for $N>0$, which
for $\alpha_s(k^2) \approx 0.25$ is at $N\approx
\alpha_s(k^2)$, leading to a splitting function $\sim
x^{-\alpha_s(k^2)}$. However, the numerical $\Gamma(N,\alpha_s(k^2))$ 
is in very good agreement with the transformation
of $p^0_{gg}(x,\alpha_s^{eff})$ until one gets extremely near the
pole -- the replacement        
$\ln(\tilde k^2/\Lambda^2) = \ln(k^2/\Lambda^2) +
3.6(\alpha_s(k^2)\ln(1/x))^{1/2}$ is extremely effective until very
small $x$ indeed (see [10] for details). 
At $x\sim 10^{-8}$ the corrections from ${\cal
O}((\beta_0\alpha_s(k^2))^2)$ and beyond act to stop the decrease of
the coupling at even smaller $x$ and freeze it at
$\alpha_s^{eff,0}(k^2)\sim 0.4\alpha_s(k^2)$.    
       
The validity of using the effective coupling can also be checked
by solving the NLO BFKL equation with running coupling, first defined in 
[11]. Now, as well as
$\beta_0$-dependent corrections to the naive LO result there is also an 
${\cal O}(\alpha_s)$ 
correction not involved with the
running of the coupling, and ignoring $\beta_0$-dependent effects 
the NLO correction to the intercept of the splitting
function is indeed $-6.5\bar\alpha_s(k^2)$ the LO. However, one may use the
same type of prescription for fixing the scale in the NLO contribution
to the splitting function as at LO, finding that, although it is 
not necessary, the $\alpha_s^{eff}$ appropriate for this NLO
correction is the same as at LO [8]. This is both for the  
$\ln(\tilde k^2/\Lambda^2) =\ln(k^2/\Lambda^2) +
3.6(\alpha_s(k^2)\ln(1/x))^{1/2}$ behaviour for $x> 10^{-8}$ and the 
freezing below this -- the $x\to 0$ NLO corrected splitting function
does not behave like the naive
$x^{-2.8\bar\alpha_s(k^2)(1-6.5\bar\alpha_s(k^2))}$, but like
$x^{-\bar\alpha_s(k^2)(1-2.3\bar\alpha_s(k^2))}$ for $\alpha_s(k^2)\approx
0.25$ [10]. Hence, the perturbative expansion is more stable.
Exact calculations at finite $x$ show that the NLO correction to the
splitting function leads to fairly small corrections to the evolution [8].
I note that other authors have considered partial resummations of the
BFKL kernel which provide improved stability besides that associated
with the running coupling [12,11,9]. These are necessary for single-scale
processes, but I feel they are less important for structure functions 
than running coupling effects.   

\section{Phenomenology.}

In order to apply these results to a study of structure functions it
is necessary to calculate the small $x$ expansions for the physical
splitting functions [13] which give the evolution of $F_2(x,Q^2)$ and
$F_L(x,Q^2)$ in terms of each other, in order to avoid factorization
scheme ambiguities. This results in $\alpha_s^{eff}$ of exactly the
same form as above. In order to compare with data one
can then combine these small $x$ expansions with the normal 
expansions in powers of $\alpha_s$ in the manner in [14]. At present
there is only sufficient information to work at LO in this combined
expansion. However, it is possible to use effective
scales at large $x$ using similar considerations as above, finding a
coupling which grows as $x\to 1$.     

A global analysis of structure function data can be performed (with
constraints applied for other data, e.g. prompt photon,
high $E_T$ jets), and the results are very good. The $\chi^2$ for 1330
data points is 1339 compared to 1511 for the standard NLO in
$\alpha_s$ MRST fit [15]. Not only is the fit improved but, whereas the
input gluon (and therefore input $F_L(x,Q^2)$) in the conventional
approach is valencelike at $Q^2\sim 1 {\hbox{\rm GeV}}^2$,
$F_L(x,Q^2)$ in this approach is the same general shape
as $F_2(x,Q^2)$ at this low $Q^2$. The prediction for $F_L(x,Q^2)$ in
the HERA range is smaller than that using the conventional approach,
being similar to [14]. 

Hence, I believe that the correct way to take
account of the $\ln(1/x)$ terms in the calculation of
structure functions is to use the combined expansion for physical 
splitting functions, and the $\alpha^{eff}$ given by
resumming $\beta_0$-dependent terms. This is preferred by
current $F_2(x,Q^2)$ data, and predictions for other quantities are
different from the conventional approach.

\end{document}